\documentclass{emulateapj}
\usepackage{color}
\usepackage{graphicx}
\newcommand{\ha}{H$\alpha$}
\newcommand{\lha}{L(H$\alpha$)}
\newcommand{\sumlha}{$\rm \Sigma L(H\alpha)$}

\newcommand{\smy}{$\rm {\it h}_{100}^{-2}~ M_\odot ~ yr^{-1}$}

\newcommand{\ca}{CL~1040$-$1155}
\newcommand{\cb}{CL~1054$-$1245}
\newcommand{\cc}{CL~1216$-$1201}

\newcommand{\edi}{EDisCS}
\newcommand{\rtwo}{$\rm R_{200}$}

\shorttitle{Mass and Redshift Dependence of SF in Clusters}
\shortauthors{Finn et al.}
\begin{document}
\title{Mass and Redshift Dependence of Star Formation in Relaxed Galaxy Clusters}

\author{Rose A. Finn\altaffilmark{1,2}, Michael L. Balogh\altaffilmark{3}, Dennis Zaritsky\altaffilmark{4}, Christopher J. Miller\altaffilmark{5},  Robert C. Nichol\altaffilmark{6}}

\altaffiltext{1}{Department of Physics, Siena College, 515 Loudon Rd, Loudonville, NY  12211 Email: rfinn@siena.edu}
\altaffiltext{2}{NSF Astronomy and Astrophysics Postdoctoral Fellow}
\altaffiltext{3}{Department of Physics, University of Waterloo, Waterloo, Canada N2L 3GI Email: mbalogh@uwaterloo.ca}
\altaffiltext{4}{Steward Observatory, 933 N. Cherry Ave., University of Arizona, Tucson, AZ  85721 Email: dzaritsky@as.arizona.edu}
\altaffiltext{5}{NOAO Cerro Tololo Inter-American Observatory, 950 North Cherry Avenue, Tucson, AZ  85719, Email: cmiller@noao.edu}
\altaffiltext{6}{Institute of Cosmology and Gravitation, University of Portsmouth, Portsmouth, UK, P01 2EG Email: bob.nichol@port.ac.uk}

\begin{abstract}
We investigate the star-formation properties of dynamically relaxed galaxy clusters 
as a function of cluster mass for 308 low-redshift clusters drawn from
the Sloan Digital Sky Survey (SDSS) C4 cluster catalog.  
It is important to establish if cluster star-formation properties have a mass 
dependence before comparing clusters at different epochs, and here
we use cluster velocity dispersion, $\sigma$, 
as a measure of cluster mass.
We select clusters with no significant substructure, a subset of the full C4 sample, so
that velocity dispersion is an accurate tracer of cluster mass.  
We find that the total stellar mass, the number of star-forming
galaxies, and total star-formation rate  scale linearly
with the number of member galaxies, with no residual dependence on cluster
velocity dispersion.
With the mass-dependence of cluster star-formation rates established, 
we compare the SDSS clusters with a sample of $z\simeq 0.75$ clusters from 
the literature and find that on
average (correcting for the mass growth of clusters between the two redshifts)
the total \ha \ luminosity of the high-redshift clusters 
is 10 times greater than that of the low-redshift clusters. This can be
explained by a decline in the \ha \ luminosities of individual cluster 
galaxies by a factor of up to $\sim 10$ since $z\simeq 0.75$.  
The magnitude of this evolution is comparable to that of field galaxies 
over a similar redshift interval, and thus the effect of the cluster environment on
the evolution of star-forming galaxies is at most modest.  
Our results suggest that the physical mechanism driving the evolution of cluster star-formation
rates is independent
of cluster mass, at least for clusters with velocity dispersion greater than $\rm 450~km~s^{-1}$, 
and operates over a fairly long timescale such that the star-formation rates of individual galaxies decline
by an order of magnitude over $\sim$7 billion years.
\end{abstract}


\section{INTRODUCTION}
Although the inverse
correlation between star-formation rates and galaxy density 
is well-established 
(e.g. {Balogh} {et~al.} 1997; {Lewis} {et~al.} 2002; {G{\' o}mez} {et~al.} 2003),
we do not yet understand whether this is due to the advanced evolution
in overdense regions or to a direct physical effect on the star formation
capability of galaxies in dense environments. The apparent rapid
evolution of galaxies in dense environments 
({Butcher} \& {Oemler} 1984; {Dressler} {et~al.} 1997; {Poggianti} {et~al.} 1999)
suggests that it may be possible to observe the quenching of 
star formation in dense regions by exploring clusters over a modest
range of redshifts ($0 < z < 1$), though great care must be taken to ensure a
fair comparison of clusters (e.g. {Nakata} {et~al.} 2005; {Andreon} {et~al.} 2006).
This evolution may be partly related to an apparent decrease in the
spiral population since $z \sim 1$ and a corresponding increase in the
S0 population (e.g. {Dressler} {et~al.} 1999; {Smith} {et~al.} 2005; {Postman} {et~al.} 2005; {Moran} {et~al.} 2007), yet the cause of the
morphology and SFR evolution remains unclear.  

We have undertaken a program
to directly measure star-formation rates (SFRs) of cluster 
galaxies over this redshift
range, and so far we have been able to demonstrate that 
star-formation rates of cluster galaxies depend on both redshift and cluster
mass ({Finn} {et~al.} 2004, 2005; {Poggianti} {et~al.} 2006).  However, we have not been able to distinguish
mass and evolutionary effects
because the high-redshift clusters typically have lower masses than the
low-redshift clusters.
Furthermore, even clusters of comparable mass and redshift 
exhibit large scatter in their star-formation properties
(possibly suggesting that another property of clusters, such as
the intracluster medium density, may play a role; {Popesso} {et~al.} 2007; {Moran} {et~al.} 2007).  
Thus, we need to have large samples of clusters with adequate mass
coverage to reliably characterize cluster SFR properties. 
The large and uniform cluster catalogs selected from the Sloan Digital Sky Survey
(SDSS; {York} {et~al.} 2000) are the best available for investigating the dependence of star-formation
rates on cluster mass.

Several groups have studied the environmental dependence of SDSS galaxy 
properties such as morphology, color, and star-formation rate.  Often,
environment is characterized in terms of the local density
(e.g. {G{\' o}mez} {et~al.} 2003; {Blanton} {et~al.} 2005; {Baldry} {et~al.} 2006), and it is generally found
that the correlations with local density do not depend strongly on the
larger-scale environment (e.g. {Balogh} {et~al.} 2004; {Blanton} {et~al.} 2005).  An alternative approach is
to identify bound systems and look for correlations with system mass,
where the latter property is measured from galaxy dynamics
(e.g. {Goto} 2005; {Rines} {et~al.} 2005), X-ray properties (e.g. {Popesso} {et~al.} 2007) or via
direct comparison to model predictions ({Weinmann} {et~al.} 2006a).  Most of
these studies have focused on trying to disentangle the effects of
galaxy stellar mass from those of environment
(e.g. {Kauffmann} {et~al.} 2004; {Weinmann} {et~al.} 2006a; {Baldry} {et~al.} 2006).  In the present paper
our goal is to understand how the total H$\alpha$ luminosity in
clusters, integrated over all bright galaxies, depends on cluster
velocity dispersion.  This allows us to make a direct comparison with
higher-redshift studies, and therefore begin to construct a
star-formation history for galaxies in clusters analogous to what has
been done in the field (e.g. {Madau} {et~al.} 1998; {Hopkins} 2004).  Some attempt
at this has already been made ({Poggianti} {et~al.} 2006; {Nakata} {et~al.} 2005); here we
expand on these works, using a larger sample of local clusters with
reliable velocity dispersions and \ha\ emission-line measurements.

Specifically, we investigate star-formation as traced by H$\alpha$ in
308 low-redshift clusters drawn from the C4 cluster catalog ({Miller} {et~al.} 2005).
We describe the cluster sample selection in \S\ref{sample} and the
selection of H$\alpha$-emitting galaxies 
in \S\ref{starforming}.  We quantify the mass-dependence of cluster
star-formation rates in \S\ref{results}, and use these results to quantify the 
environment-dependent evolution by comparing with a higher-redshift
sample of clusters in \S\ref{evolution}.
We summarize the main results and present conclusions in \S\ref{conclusions}.
We adopt a $\Lambda$CDM cosmology, 
assuming $\Omega_0 = 0.3$, $\Omega_\Lambda = 0.7$, and $\rm {H_0} = {70~km~s^{-1}~Mpc^{-1}}$ unless
otherwise noted.

\section{CLUSTER SAMPLE \& MEMBER GALAXIES}
\label{sample}
\subsection{Sample Selection \label{selection}}
We select our cluster sample from the C4 DR5 cluster catalog 
({Miller} {et~al.} 2005).
We apply a minimum redshift cut of $z = 0.05$ 
to minimize aperture effects of the fiber spectra ({Kewley} {et~al.} 2005) and a 
maximum redshift cut of 0.09 to
minimize incompleteness of the galaxy catalog due to the spectroscopic magnitude
limit.  
This leaves a sample of 923 clusters.

The virialized region of a cluster scales with cluster mass,
so we need a reliable mass estimate in order to properly 
select member galaxies for each cluster. 
There are many ways to estimate cluster mass.
For example, {Weinmann} {et~al.} (2006b) estimate halo mass using halo occupation models, and 
{Rines} \& {Diaferio} (2006) 
use the distribution of galaxies in $\Delta v - \Delta r$ space to derive cluster mass
profiles.  However, the simplest method, and the one most easily
applied to higher redshift samples, is to use the cluster velocity
dispersion.  Velocity dispersion can be an accurate tracer of cluster mass for relaxed
systems that are not dominated by substructure, and for which there are
a sufficiently large number of redshifts available
(e.g., {Miller} {et~al.} 2005).  Furthermore, 
velocity dispersion can be directly related to other cluster properties
such as virial radius using the virial theorem and 
cluster size in the redshift dimension assuming an underlying
velocity distribution.  

\begin{figure*}[h]
\plotone{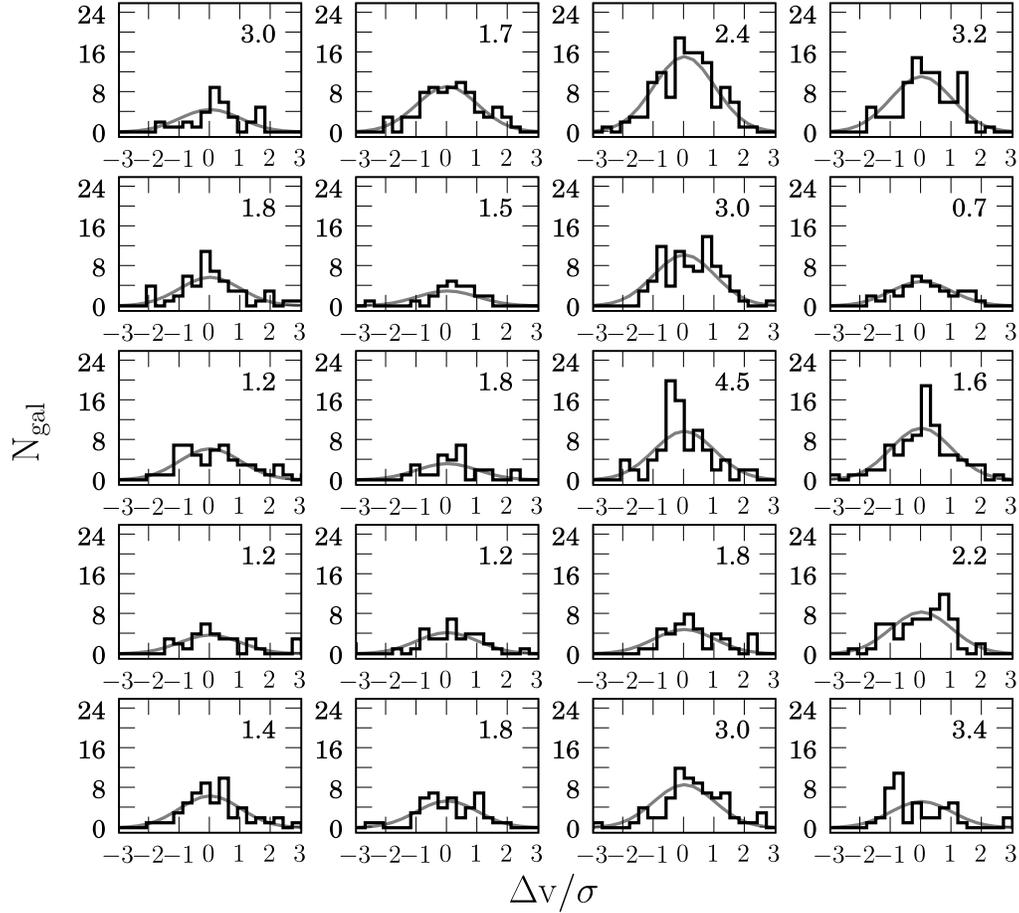}
\caption{Velocity histograms for first 20 clusters.  
  The gray curve shows a gaussian with
width equal to the cluster velocity dispersion and normalized so that the area 
within $2\sigma$ matches the observed number of galaxies in the same velocity
slice.  The number in the upper right of each panel is the value of $\chi^2_\nu$.  
Clusters with $\chi^2_\nu > 2.0$ are rejected due to the presence
of significant substructure within or near the cluster.\label{velhist}}
\end{figure*}

We will therefore 
use velocity dispersion as a measure of cluster mass.  However, for
this approach to be valid a minimum requirement is that the clusters be in
approximate virial equilibrium, with a well-defined measurement of
velocity dispersion.  To satisfy this condition, we exclude clusters
which do not appear to be dynamically relaxed.  
This selection may reduce the scatter in cluster properties
at fixed mass compared with an unbiased sample.  It may also
bias the average star-formation rates, if
cluster mergers directly affect star formation in member galaxies.
However the advantage is that we can more directly identify 
any correlations with system mass, independently of dynamical effects.
Clusters with no significant substructure 
should have velocity distributions that are nearly gaussian,
and we identify unrelaxed clusters by
quantifying how much their velocity distribution 
deviates from this.  For each cluster, we fit a gaussian with a width
equal to the cluster velocity dispersion from {Miller} {et~al.} (2005).  We normalize the 
gaussian so that the area within 
$\Delta v < 2\sigma$ matches the observed number of cluster 
members in that velocity range.  
To illustrate, we show the velocity distributions and gaussian
fits for the first twenty clusters in Figure \ref{velhist}.  

We use the $\chi^2_\nu$ statistic to quantify the 
deviation of the velocity distribution from gaussian, where
\begin{equation}
\label{rtwo}
\chi^2_\nu = \frac{1}{(N_{bin}-3)} \sum_{i=1}^{N_{bin}} \frac{(N_{obs}-N_{gauss})^2}{\sigma_{N_{obs}}^2} .
\end{equation}
$N_{bin}$ is the number of bins, $N_{obs}$ is the number 
of galaxies observed in each bin, $N_{gauss}$ is the 
number of galaxies expected in a gaussian distribution, and $\sigma_{N_{obs}} = \sqrt{N_{obs}}$
is the error associated with the expected counts. When $N_{obs}$ is zero,
we set $\sigma_{N_{obs}}$ to 1.
We use a bin size of 0.3 times the velocity dispersion.  
After visual inspection of the velocity histograms, we find that 
$\chi_\nu^2 < 2.0$ excludes clusters with significant 
substructure within the cluster itself or in the nearby environment.
If a cluster does not make the $\chi^2_\nu$ cut, we try recentering the
velocity distribution and then recalculate $\chi_\nu^2$.  If $\chi_\nu^2$ 
of the recentered distribution meets the criteria, 
we keep the cluster and the new central velocity.  

The $\chi^2_\nu$ criteria is not uniform with velocity dispersion; 
$\chi_\nu^2$ is systematically lower 
for clusters with lower numbers of galaxies.  
Therefore, we impose a minimum velocity 
dispersion of $\rm 450~km~s^{-1}$
because the $\chi_\nu^2$ cut becomes unreliable below this point, leaving 
a sample of 452 clusters.  
In addition, we note that at all values of velocity dispersion 
this cut will preferentially select clusters 
with fewer galaxies, and
we must keep this selection effect in mind when interpreting our results.

\begin{figure}[h]
\plotone{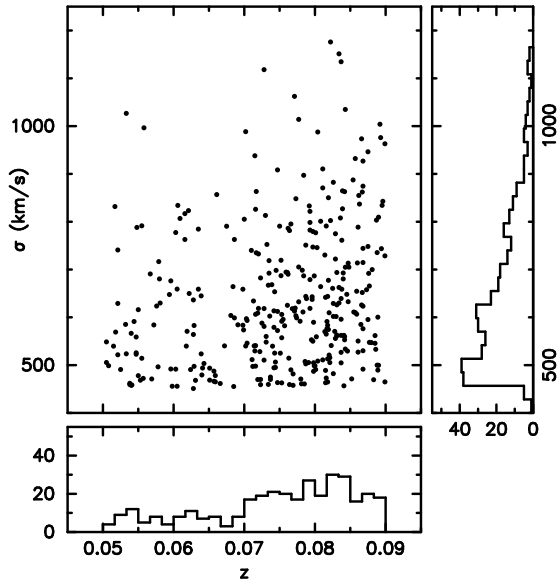}
\caption{Cluster velocity dispersion versus redshift.  
The bottom panel shows a histogram of number of clusters versus redshift.  The side 
panel shows the histogram of number of clusters versus velocity dispersion.
\label{sigmaz}}
\end{figure}

We apply the $\chi^2_\nu < 2.0$ criteria to the remaining
clusters with $\rm \sigma > 450~km~s^{-1}$, which leaves 308 clusters.  
We show velocity dispersion
versus redshift for the final sample of clusters 
in Figure \ref{sigmaz}.  
There is no trend in velocity dispersion versus redshift for the final sample.

\subsection{Cluster Membership}
The size of the virialized region of a cluster scales with cluster mass.
Therefore, 
our radial cut for selecting cluster members must scale with cluster mass
to sample analogous regions of each cluster.
We characterize the radial extent of each cluster in terms of 
\rtwo, which is the
radius inside which the enclosed density is 200 times the critical
density and approximates the virial radius of the cluster.  
Using the redshift dependence of the critical density and
the virial mass to relate line-of-sight velocity dispersion, $\sigma_x$,
to cluster mass, we express \rtwo \ as:
\begin{equation}\label{eqn-r200}
R_{200} = \sqrt{\frac{9}{800 \ \pi \ G \ \rho_c(z)}} \ \sigma_x,
\end{equation}
where the critical density, $\rho_c(z)$, is given by the following equation:
\begin{equation}
\rho_c(z) = \frac{3\ H_0^2}{8 \ \pi \ G} \ (\Omega_\Lambda + \Omega_0 (1 + z)^3).
\end{equation}
Simplifying, we can write \rtwo \ as
\begin{equation}
R_{200} = 2.02 \ \frac{\sigma_x}{1000~{\rm km/s}} \ \frac{1}{
\sqrt{\Omega_\Lambda + \Omega_0 (1+z)^3}}\  h_{70}^{-1} \ {\rm Mpc}.
\label{rtwoeqn}
\end{equation}
In deriving Equation~\ref{eqn-r200} we have implicitly assumed that galaxies are orbiting
isotropically in a single isothermal sphere potential, so $\sigma_x$ is
related to the circular velocity $V_c$ by $\sigma_x=V_c/\sqrt{2}$.

\begin{figure}[h]
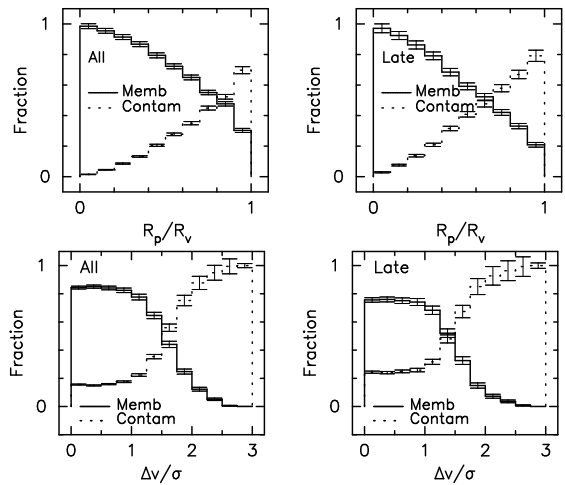

\plotone{f3a.ps}
\plotone{f3b.ps}
\caption{The fraction of members (solid curve) and 
non-members (dashed line) per bin as a function of projected cluster-centric
radius, $\rm R_p$, normalized by the virial radius (top panels) and 
velocity cut, $\Delta v$, normalized by the cluster velocity dispersion (bottom panels) 
in the {Yang} {et~al.} (2004) model. 
The left panels show all galaxies and the right panels show
late-type galaxies only.  The errorbars show the $1-\sigma$ Poisson noise associated with
the galaxy number counts.  
\label{histdr}}
\end{figure}

We use the mock galaxy redshift survey of {Yang} {et~al.} (2004) to help define
the optimal selection of
cluster members in velocity space.  We use the 300~$h^{-1}$~Mpc simulation and 
select all halos with masses greater than 
$\rm 10^{14}~M_\odot$ so that the minimum mass of the mock clusters is
consistent with the mass threshold imposed on the SDSS clusters 
by our $\rm \sigma > 450~km~s^{-1}$ cut.  This results in 1062 virialized clusters
containing a total of a half million galaxies.  The velocity distributions of 
the mock clusters are defined to be gaussian.  Thus the mock clusters should be comparable
to our cluster sample because we exclude clusters whose velocity distributions
deviate significantly from gaussian.  We then 
select galaxies around each halo that have $M_{B_J} < -18$ and velocities
within $\pm 3\sigma$ of the central velocity.  
In the top panels of Figure \ref{histdr}, we show the fraction of observed galaxies that are members
(solid line) and non-members (dotted line) in each radial bin as a function of 
projected distance from the cluster center, where members are those galaxies 
physically located within the virial radius 
of the cluster.  Yang {et al.} define the virial
radius using $\rm R_{180}$;  their definition of $R_{vir}$ is thus 5\% larger
than our value of \rtwo, but this slightly larger radial cut does not significantly affect
the results.
The left panel of Figure \ref{histdr}
shows completeness and contamination for all galaxy types, and the number of
contaminating galaxies exceeds the number of member galaxies at a radius of 
$0.9$ times the virial radius.  The right panel shows completeness and contamination for 
late-type galaxies only, with the number of contaminating late-type galaxies exceeding the 
number of member late-type galaxies at a projected radius of 0.7 times the 
virial radius.  
In the bottom panel of Figure \ref{histdr}, we show the fraction of members and
non-members per bin as a function of velocity cut, including all galaxies  
with a projected separation less than $R_{vir}$.  
For a gaussian distribution of velocities, 
a $2\sigma$ membership cut within the virial radius provides fairly high 
completeness ($\ge$95\%), while increasing the velocity cut to $3\sigma$ adds
more non-members than members.
When working within \rtwo, we therefore limit our analysis to galaxies within 
$2\sigma$.  Note that the results are not sensitive to the magnitude cut; we find 
no significant difference for magnitude cuts ranging from $M_{B_J} < -16$ to 
$M_{B_J} < -19$.  Thus, while a magnitude cut of $M_{B_J}=-19$ is more comparable to the
$r$-band magnitude cut we impose (as discussed below), the $M_{B_J}=-18$ allows us 
to increase signal-to-noise without affecting the results.  

Finally, we apply an absolute magnitude cut of $M_R \le -20.68$ to
the cluster galaxies.  This corresponds to
the spectroscopic completeness limit of $r = 17.77$ at $z = 0.09$, the
maximum redshift of our cluster sample.  The absolute magnitude cut
takes into account the maximum $k$-correction possible
({Blanton} {et~al.} 2003a).  
Note that this is only about $0.5$ magnitudes
fainter than the characteristic luminosity $M^\ast$ ({Blanton} {et~al.} 2003a).
Our analysis is therefore restricted to relatively bright galaxies, and
it should be kept in mind that star formation activity can be
substantially higher for less massive galaxies (e.g. {Baldry} {et~al.} 2006).

\subsection{Control Sample}
We construct a control (field) sample to compare with our cluster
measurements and to check for systematic effects in 
our analysis.  
For each cluster, including those with substructure, we center on a random galaxy 
within the DR5 redshift survey area, {\em regardless of its magnitude}, and we 
avoid centering near clusters.  We then
select surrounding galaxies with magnitudes brighter than
$M_R = -20.68$ that lie within $R_{200}$ and $\Delta v \le 2 \sigma$.
We select five such randomly positioned samples for each cluster, resulting in a control sample of 2260 independent
pointings.  
We generate control samples for clusters that meet the redshift cuts and 
lower velocity dispersion limit cut.  However, we do not 
limit the control sample based on the substructure of its parent cluster.  
As a result, the control sample spans 
a slightly different range of velocity dispersion than the cluster sample.

\subsection{Correction for Incomplete Fiber Sampling\label{specsampling}}
Not all cluster members are targeted by SDSS spectroscopy, so we must 
correct the cluster and control samples 
for incomplete spectroscopic sampling when calculating 
integrated cluster
properties such as the number of galaxies, total stellar mass, and total star-formation rate.
To estimate the completeness of the spectroscopic sampling, we count the number of galaxies
in the photometric catalog with an r-band magnitude greater than that corresponding to 
$M_r = -20.68$ at the cluster redshift.  We compare this to the number of galaxies 
with spectra in the same region and magnitude slice, and multiply the integrated cluster 
properties by the ratio of photometric to spectroscopic sources.
For the cluster sample, the average completeness is $0.86 \pm 0.15$, $0.86 \pm 0.11$, and 
$0.87 \pm 0.10$ for radial cuts at 0.5, 1.0, and $2.0\times$\rtwo.  The median completeness is
0.89.  For the control sample, the average completeness within 0.5, 1, and $2\times$\rtwo \ 
is $0.92 \pm 0.20$,$0.89 \pm 0.16$, and $0.88 \pm 0.12$.  The completeness estimates 
for the control fields are in good agreement with the expected spectroscopic
completeness of 90\% ({Blanton} {et~al.} 2003b), while the completeness estimates for the clusters are slightly
lower due to the higher density of sources. 

\section{Star-Forming Galaxies}
\label{starforming}
\subsection{Definition of Star-Forming Galaxies}

We require star-forming galaxies to have
\ha \ luminosities greater than
$\rm 0.41 \times 10^{40} \ ergs \ s^{-1}$
to sample the same fraction of the 
\ha \ luminosity function in each cluster.
This limit corresponds to the minimum \ha \ flux detected at $z = 0.09$, 
$\rm f_{H_\alpha} = 2.0 \times 10^{-16} \ \rm ergs \ s^{-1}\ cm^{-2}$, 
which we determine empirically.  This luminosity limit corresponds
to a SFR of 0.08~\smy \ if we assume a typical extinction value of 1 magnitude 
at \ha\ ({Kennicutt} {et~al.} 1994).

We also require star-forming galaxies to have a rest-frame H$\alpha$
equivalent width greater than 4\AA, which allows us to minimize
contamination from uncertainties on the line width measurements and
stellar absorption of inactive galaxies (e.g. {Balogh} {et~al.} 2004).  
However, imposing this cut necessarily means we will underestimate the
total amount of star formation in an H$\alpha$ flux-limited sample.  To
test this, we consider the subsample of galaxies with $z<0.075$, where
we can reliably push to lower EW limits.  We find that at most 5\% of
the total star formation in our flux-limited sample occurs at EW$<4$\AA.
As this small contribution is somewhat uncertain (since measurement
uncertainties and stellar absorption still play a role) we do not
correct for it.

We exclude AGN from the sample of star-forming 
galaxies using the AGN classification provided by 
{Miller} {et~al.} (2003), which have been updated to include the DR5 galaxies.  
We find $22\pm 13$\% of galaxies by number with significant
AGN emission within $\Delta v < \pm 2 \sigma$ and $1\times$\rtwo, 
and they contribute an average(median) of $21\pm 13$\%(29\%) of the total \ha \ flux.
This result is consistent with the AGN 
fraction measured by {Miller} {et~al.} (2003) and {Kauffmann} {et~al.} (2003a).

\subsection{Star-Formation Rates \& Stellar Mass \label{sfrconv}}
To calculate SFR from observed \ha \ flux, we first apply an 
aperture correction which is computed by comparing the flux within the 
$R$-band petrosian radius with the flux in the fiber radius ({Balogh} {et~al.} 2004).
We then convert \ha \ flux to luminosity by multiplying by
$4 \pi d_L^2$, where $d_L$ is the luminosity distance.
We use the Kennicutt star-formation relation ({Kennicutt} {et~al.} 1994) 
to convert \ha \ luminosity to SFR, where
\begin{equation}
\rm
1~ erg \ s^{-1} = 7.9 \times 10^{-42} \ M_\odot \ yr^{-1}.
\end{equation}

We correct for extinction using values of A$_z$ from {Kauffmann} {et~al.} (2003b), 
assuming the extinction law is of the form 
$\tau_\lambda \propto \lambda^{-0.7}$ and thus $A_r = A_z + 0.23$.
The {Kauffmann} {et~al.}
sample has been updated to include DR4 galaxies but not DR5.  
Therefore, 23\% of the galaxies do not have an extinction value from
these catalogs.  To estimate the extinction for these galaxies,  
we use the DR4 galaxies to define a relationship 
between A$_r$ and M$_r$.  
The median value of A$_r$ for galaxies brighter
than our magnitude cut (M$_r < -20.68$) 
remains fairly constant at a value of 0.76 magnitudes.  
Therefore, we assume 0.76 magnitudes of extinction 
at \ha \ for the galaxies not included in the {Kauffmann} {et~al.} (2003b) catalog.
The standard deviation in A$_r$ about the mean is 0.30 magnitudes, and changing
the extinction value by one standard deviation alters the inferred star-formation
rates by 30\%.  Thus, the error associated with the extinction correction
is small compared to other sources of error and does not significantly affect our results.

\begin{figure}
\plotone{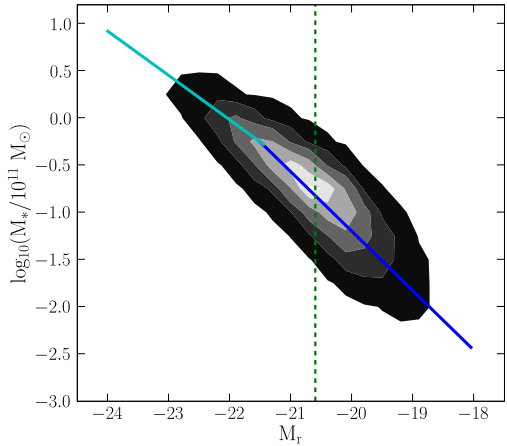}
\caption{Distribution of stellar mass versus $M_r$ for the DR4 galaxies.  
Blue and cyan line show average relation used
to estimate stellar mass from $M_r$ for galaxies not included in the 
{Kauffmann} {et~al.} (2003b) catalogs. The dashed green vertical line shows the magnitude cut used when selecting cluster
members.\label{stellarmr}}
\end{figure}

Stellar mass is the integral of a galaxy's star-formation history and enables us to study
past star-formation efficiency as a function of cluster mass.  
Again, we use stellar mass estimates from {Kauffmann} {et~al.} (2003b).
To estimate the stellar mass for galaxies not included in the Kauffmann 
{et~al.} catalog, 
we use the DR4 galaxies to define a relation between M$_r$ and stellar mass. 
We show the result in Figure \ref{stellarmr}, where the blue line shows the relationship
we use for
galaxies with M$_r > -21.5$ (blue line) and 
the cyan line shows the slightly flatter relationship for galaxies with M$_r < -21.5$.

\section{Mass Dependence of Cluster SF Properties}
\label{results}
\begin{figure}[h]
\plotone{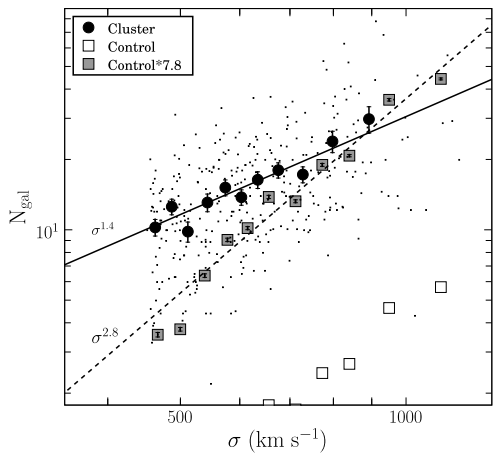}
\caption{The total number of galaxies as a function of cluster velocity dispersion for all galaxies 
with velocities 
within $\pm 2 \sigma$ and projected cluster-centric distance less than $1\times R_{200}$. 
The small black points show the values for individual clusters, and
the larger symbols show the average in equally populated bins for the cluster (black circles) and 
control (open squares) samples.  Errorbars 
show the error in the mean in each bin.
The filled squares show the control data, scaled so that the average counts
in the control fields equal the average counts in the cluster fields.  
The solid line shows the best power-law fit for the cluster data, and the
dashed line shows the best power-law fit for the control data.
\label{ngalsigma}}
\end{figure}

We first investigate how the number of cluster members brighter than
$M_r=-20.68$ depends on cluster velocity dispersion.
In Figure \ref{ngalsigma}, we show the number
of galaxies within \rtwo \ and with $\Delta v < 2 \sigma$ for the 
cluster (black circles) and control fields (open squares).  
The small black points show the values for the individual clusters.
The number of galaxies is strongly correlated with velocity dispersion for both 
samples.   Note that the number of galaxies in the control sample
increases as approximately $\sigma^3$ as expected since the cylindrical
selection volume scales in this way.  The
cluster fields are overdense relative to the field, by a factor of
6---23 that appears to depend on velocity dispersion.  Note that the
clusters are not corrected for residual field contamination; this
figure demonstrates that we are overestimating the number of cluster
galaxies by no more than about 10 per cent due to this effect.
The filled squares show the control data rescaled by a factor of $7.8$,
to allow easy comparison of the slope of the 
cluster and control samples.

It is evident that
the number of galaxies in the cluster fields follows a 
flatter trend versus velocity dispersion 
than the number of galaxies in the control fields.  
However, we note that there are several important biases here that must be
accounted for.  First, the uncertainties on both values are correlated, 
such that an overestimated $\sigma$ leads to a larger volume and, hence, 
an overestimated $N_{\rm gal}$.  For the field sample, the uniform galaxy 
density means one expects $N_{\rm gal} \propto \sigma^3$, so 
$dN_{\rm gal} \propto \sigma^2 \ d\sigma$.
However, for the cluster population, the 
galaxy overdensity in a given cluster is expected to decline as
approximately $R^{-2}$, so the total number of galaxies within
$R_{200}$ (which includes the cluster and surrounding field population)
should scale like $N_{\rm gal} \propto \sigma^\alpha$, with
$\alpha<3$.  At the extreme, where the background population is
negligible and most of the cluster population is well within $R_{200}$,
$N_{\rm gal}$ is independent of uncertainties in $\sigma$.
For these reasons, at a fixed value of $\sigma$, observational
uncertainties alone will scatter the cluster measurments of 
$N_{\rm gal}$ along a slope that is shallower than that of the field
measurements.  We do not attempt to quantify this however, as it is
sensitive to the radial distribution of galaxies within clusters, and
the relative contribution of foreground and background galaxies.
Finally, cluster selection biases will play a role.  The
imposed limit on $\sigma$ results in a bias at low $\sigma$ toward
clusters where $\sigma$ is overestimated.  Furthermore, our
$\chi^2$ cut on velocity distribution tends to favor systems with fewer
members, which will result in a bias at high $\sigma$ toward fewer
$\rm N_{\rm gal}$.  For all these reasons we do
not attribute much significance to the apparently flatter slope of the
clusters.  
Hereafter, when attempting to quantify cluster star-formation
properties as a function of $\sigma$, 
we will normalize by the total number of member galaxies so that we reduce the $\sigma$-dependence
of the dependent variable.  Similar results would be obtained if we
normalized by the total stellar mass (we will show below that the two
are strongly correlated, as expected); however for comparison with
high-redshift studies it is preferable to use a simple measurement like
$\rm N_{\rm gal}$ if possible.

\subsection{Distribution of \ha \ Luminosities}
\begin{figure}[h]
\plotone{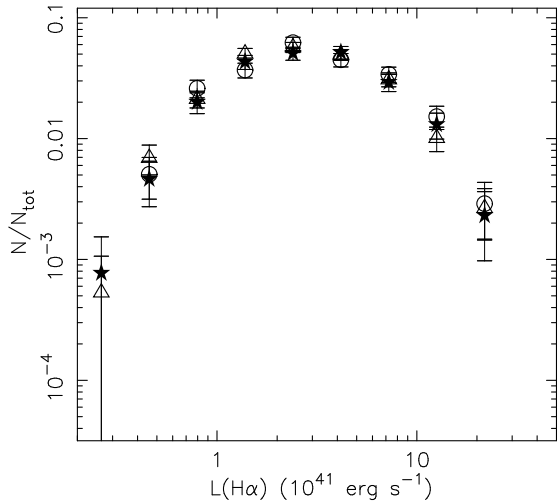}
\caption{Number of galaxies per bin divided by total number of galaxies versus \lha.  Stars, circles, and triangles show galaxies
from clusters with $\sigma < 575$, $575 \le \sigma < 700$, and $\sigma \ge 700$~km~s$^{-1}$, respectively.  Errorbars 
reflect $1-\sigma$ Poisson noise.  
\label{histlhasdss}}
\end{figure}

We show the distribution of \ha \ luminosities as a function of 
cluster velocity dispersion in Figure \ref{histlhasdss}.  
We split
the cluster sample into three groups based on mass, and dividing at $\sigma = $575~km~s$^{-1}$ and $\sigma = $700~km~s$^{-1}$ 
yields an approximately equal number of galaxies per group ($\sim 800$).  
The points (stars, open circles, open triangles) show the number of galaxies per luminosity
bin divided by the total number of galaxies in that mass division ($\sim 800$ galaxies), 
and the errorbars show the $1-\sigma$ Poisson noise.  
The three sets of points are 
consistent within the errors, showing that the distribution of \ha \ luminosities is not a strong
function of cluster velocity dispersion.

\begin{figure}[h]
\plotone{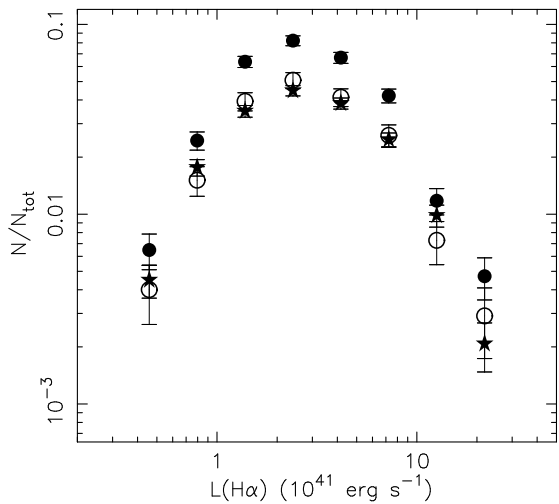}
\caption{Number of galaxies per bin divided by total number of galaxies versus \lha \ for the cluster (star) 
and control (filled circles) samples.  
The open circles show the distribution for the control fields after being scaled to meet the 
average counts of the cluster fields.  
Errorbars reflect poisson
noise.  The 
\label{allhistlha}}
\end{figure}

In Figure \ref{allhistlha}, we compare the distribution of \ha \ luminosities for the cluster (stars) 
and control (filled circles) samples.  
The open circles show the distribution for the control fields after being scaled to meet the 
average counts of the cluster fields.  The errorbars for the cluster and scaled control
fields overlap, indicating that 
the shapes of the cluster and control \ha \ luminosity distributions are similar.
This implies that the primary difference between
the cluster and control sample is in the fraction of star-forming
galaxies, with the 
star-forming fraction of the control fields a factor
of $\sim 1.6$ higher than the cluster fields.  
This is consistent with previous results based on smaller samples
({Carter} {et~al.} 2001; {Balogh} {et~al.} 2004; {Rines} {et~al.} 2005), which have shown
that the distribution of \ha \ luminosities and EWs of the star-forming population is independent of environment.  

\subsection{Integrated Cluster Properties\label{intcluster}}
\begin{figure*}[h]
\plotone{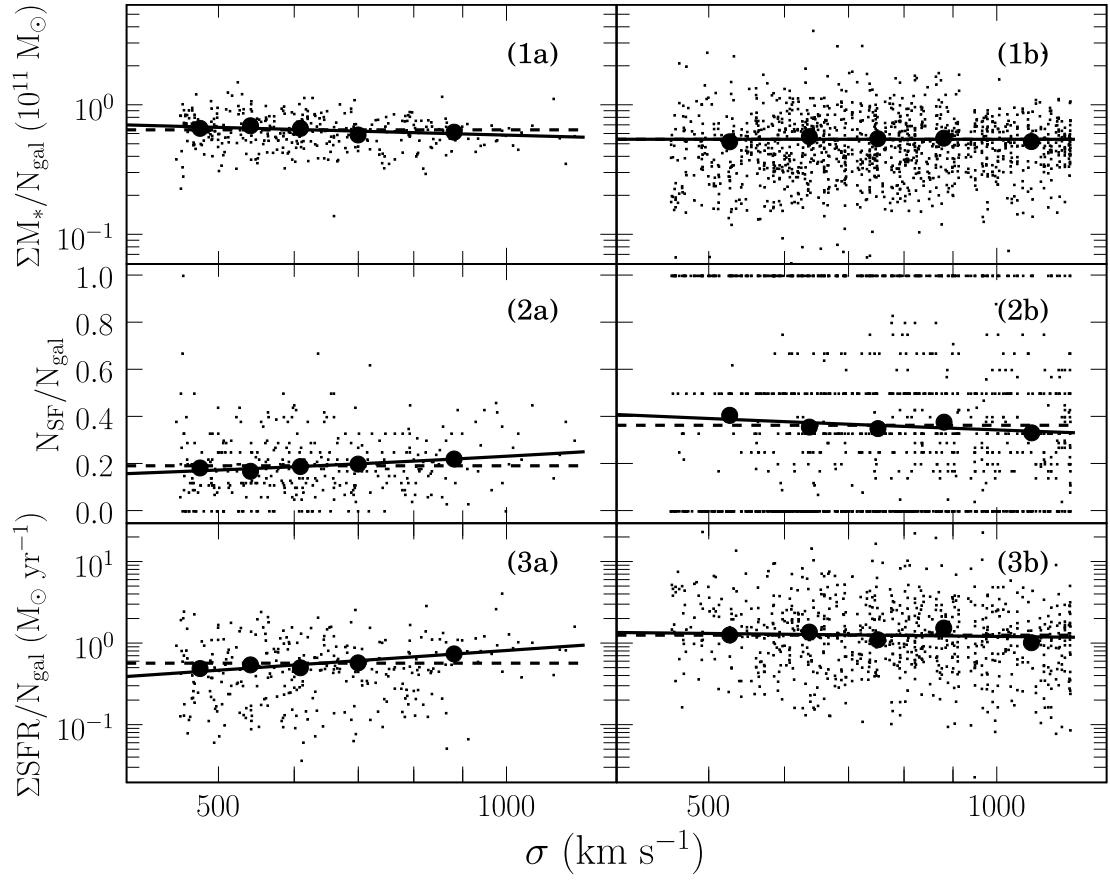}
\caption{(1) Total stellar mass, (2) number of star-forming galaxies,
and (3) the total SFR per total number of galaxies versus cluster velocity dispersion for all galaxies 
with velocities 
within $\pm 2 \sigma$ and projected cluster-centric distance less than $1\times R_{200}$. 
The left and right panels show the cluster and control data, respectively.
In each panel, the small black points show the values for individual clusters or control fields, and
the larger symbols show the average in equally populated bins.  Errorbars 
showing the error in the mean in each bin are smaller than the symbols.  
The solid line shows the best power-law fit for the data, and the
dashed line shows the average value of the y-axis variable.
\label{allvsigma}}.
\end{figure*}

We now look for a mass dependence in the integrated properties of
the star-forming cluster galaxy population.   In Figure \ref{allvsigma}, 
we show the total stellar mass (panel 1), the number of star-forming galaxies (panel 2), and the total
SFR (panel 3) 
for all galaxies within \rtwo \ and with $\Delta v < 2 \sigma$  
versus cluster velocity dispersion, where all the dependent variables have
been normalized by the number of galaxies in each cluster to reduce our
sensitivity to correlated errors.  
The left and right columns of Figure \ref{allvsigma} show the cluster
and control data, respectively.  The small black points show the values for individual fields,
and the filled circles show the average in equally populated bins.  The
errorbars, often smaller than the filled circles, show the error in the mean.
The solid line shows the best power-law fit in each panel, and the dashed horizontal line shows 
the average value to show what one would expect if there is no dependence on cluster mass.  
None of the power-law fits deviates significantly from a flat relationship,
indicating all three measures of star-formation efficiency are independent of cluster mass,
at least for clusters with $\sigma > 450$~km~s$^{-1}$.
An equivalent way to express these results is that the total stellar mass, the number of 
star-forming galaxies, and the total SFR scale linearly with the number of galaxies, independent of
cluster velocity dispersion.

Our results are generally consistent with those of other studies that
have measured star-formation efficiency as a 
function of cluster mass using smaller samples of SDSS clusters.
{Poggianti} {et~al.} (2006) use a smaller subset of the C4 clusters
and find that the fraction of star-forming galaxies decreases
as velocity dispersion increases from $\sim$300 to $\sim$500~km~s$^{-1}$ and then levels off.
We do not probe to low enough velocity dispersions to compare results 
at $\sigma = 300$~km~s$^{-1}$, but our results are consistent with a flat relationship
for $\sigma > 450$~km~s$^{-1}$. 
Note that this result is unchanged if we keep AGN in our sample (e.g. {Poggianti} {et~al.} 2006).
These results are fully consistent with the lack of mass dependence
observed from smaller samples of both optically-selected ({Goto} 2005)
and X-ray selected ({Popesso} {et~al.} 2007) clusters.

\section{Evolution of Cluster Star Formation}
 \label{evolution}
We now use our SDSS cluster sample as a low-redshift baseline and 
compare with star-formation properties of higher-redshift clusters from the
literature.  We will focus on those few studies that have comparably
complete observations of H$\alpha$ emission.

There are a growing number of \ha \ surveys of galaxy clusters at
$z > 0.1$, including both spectroscopic ({Couch} {et~al.} 2001; {Balogh} {et~al.} 2002) and narrow-band imaging 
({Balogh} \& {Morris} 2000; {Finn} {et~al.} 2004; {Kodama} {et~al.} 2004; {Finn} {et~al.} 2005) surveys, which
are suitable for our purposes.
For this comparison we use only the narrow-band H$\alpha$ imaging data for 
three $z \simeq 0.75$ clusters from 
{Finn} {et~al.} (2005) because they provide the longest redshift baseline
and the narrow-band observations for all three clusters were made in a uniform way using
the same telescope and near-infrared camera.  
In addition, the three clusters are from the ESO Distant Cluster
Survey (EDisCS; {White} {et~al.} 2005), and so they have readily available
derived-data products (such as k-corrected absolute magnitudes and estimates 
of the total number of member galaxies; {Pell\'o} {et~al.} 2007; {Rudnick} {et~al.} 2007) 
that enable a more precise comparison with the SDSS clusters.  Furthermore, 
spectroscopic (Halliday {et~al.} 2004; Milvang-Jensen {et~al.} 2008), 
weak lensing (Clowe {et~al.} 2006), and 
X-ray studies (Johnson {et~al.} 2006) show that the velocity dispersions
provide a reliable estimate of cluster mass for these clusters.  The \edi \ clusters \ca, \cb, and \cc \
are at redshifts of 0.704, 0.748 and 0.794, respectively.  
Note that we do not have the data necessary to identify AGN in the $z \simeq 0.75$ clusters.
As a result, we include SDSS galaxies with AGN emission as well, 
and hereafter we discuss \ha \ emission in terms of \lha  \ rather than SFR.

\subsection{Integrated \ha \ Luminosity \label{evoltot}}

The integral of the \ha \ luminosity function gives the total \ha \ luminosity per
cluster or integrated SFR if AGN are excluded.  This quantity is analogous to the
volume-averaged SFR that is commonly used to quantify the evolution of field galaxies
(e.g., {Madau} {et~al.} 1998; {Hippelein} {et~al.} 2003; {Hopkins} 2004), and so we use the total \ha \ luminosity 
as a means of quantifying the evolution between the SDSS and $z\simeq 0.75$
clusters.

When calculating the integrated \ha \ luminosity, we apply several selection criteria
to ensure that we are comparing complete and analogous galaxy 
samples in the higher and low-redshift clusters.  
First, we apply the same radial and velocity cuts to all clusters, 
including all galaxies within 0.5$\times$\rtwo \ and $\Delta v < 3 \sigma$.  
The radial cut matches the areal coverage
of the \edi \ clusters, and the velocity cut matches the filter width for \cc. 
The filters for \ca \ and \cb \ correspond to a velocity width closer to 6$\sigma$.  
The correction for the extra field contamination in these filters, which we detail in the Appendix, amounts to 
scaling the integrated \lha \ of \ca \ and \cb \ by
a factor of $0.81$.
Second, we apply the same rest-frame equivalent width cut of 10\AA \ to all star-forming 
galaxies, which is the minimum \ha \ equivalent width detected in the $z\simeq 0.75$ narrow-band imaging 
surveys.  
Third, we include only those galaxies with $M_r < -20.68$, where the
values of M$_r$ for the \edi \ clusters are taken from {Pell\'o} {et~al.} (2007).
Fourth, we require the SDSS galaxies to have \ha \ luminosities 
that lie above the flux 
limit of the $z\simeq 0.75$ \ha \ imaging, which we discuss in more detail in \S\ref{distlha}.  
Finally, we correct the integrated \lha \ of the SDSS clusters by the spectroscopic completeness 
as described in \S\ref{specsampling}.

\begin{figure}[h]
\plotone{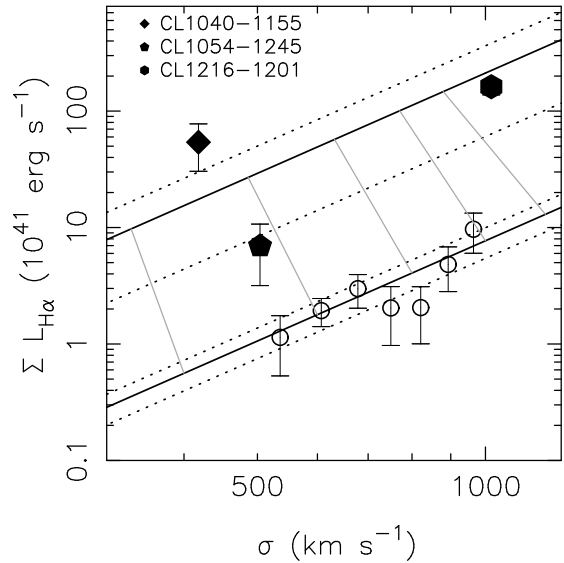}
\caption{Integrated \ha \ luminosity versus cluster velocity dispersion 
for C4 clusters (open circles) and $z\simeq 0.75$ \edi \ clusters (filled symbols).
The solid black lines show a slope of 2.9,
which is the best-fit slope derived from the SDSS clusters.  
The dotted lines show the standard deviation associated with
the zeropoint of powerlaw fits.  The solid gray lines
link the $z=0.07$ clusters with their progenitors at $z = 0.75$.
\label{sfrmembsigma3hiz}}
\end{figure}

The cluster version of the Madau plot ({Madau} {et~al.} 1998) needs to take into account that the 
integrated \ha \
luminosity scales with cluster mass.  Therefore, instead
of showing integrated \ha \ luminosity versus redshift, we show \sumlha \
versus cluster velocity dispersion in Figure \ref{sfrmembsigma3hiz} 
and measure an offset between the low and high-redshift
cluster samples.
The open circles show the average \sumlha \ in equally-populated bins versus 
velocity dispersion for the SDSS clusters,
and the errorbars show the error in the mean.  The filled points show \sumlha \ for 
each individual high-redshift cluster, and the errorbars are the
uncertainty of the sum.  
The bold solid lines in Figure \ref{sfrmembsigma3hiz} show a slope of 2.9,
which is the best-fit slope derived from the SDSS clusters.  
We assume the same slope for the $z\simeq 0.75$ clusters and fit the zeropoint 
by least-squares minimization.  The 
thin dotted lines show the standard deviation associated with zeropoint for 
both the SDSS and EDisCS power-law as determined by bootstrap resampling.
The offset between the best-fit lines for the SDSS and $z \simeq 0.75$ clusters corresponds to 
a factor of $\sim$26$\pm$16 increase in \sumlha \ when measured at a fixed velocity dispersion, with
the uncertainty dominated by the uncertainty in the high-redshift zeropoint.
Clusters evolve, however,
and a typical cluster has roughly doubled its mass since $z \simeq 0.75$.  The gray solid lines
in Figure \ref{sfrmembsigma3hiz} link the $z=0.07$ clusters with their progenitors at $z = 0.75$, where
we have used the semi-analytic models of {Wechsler} {et~al.} (2002) to estimate the redshift-evolution 
of cluster mass.  Taking the mass evolution into account, the inferred evolution for 
a given cluster is a median factor of $\sim10\pm6$, with
the amount of evolution increasing as cluster mass decreases.

To measure the average evolution in the total \ha \ luminosity 
between the SDSS and higher-redshift clusters, we
assume that the slope of the $\Sigma L(H\alpha) - \sigma$ relation
is the same for the SDSS and higher-redshift clusters.  
The change in zero point then indicates
the average evolution.  However, the slope of the relation
could also vary with redshift.  
For example, Poggianti et al. (2006) present a detailed analysis of the
star-forming population for the complete \edi \ sample of 17 
$0.4 < z < 0.8$ clusters, and their results 
suggest that the trend of
total star-formation rate versus cluster mass is flatter at higher redshift.
Poggianti et al. use the [O~II] emission line as a star-formation indicator, 
and here we limit our comparison to the 3 \edi \ clusters that have \ha \ luminosities. 
As a result of the small sample size, we 
can not independently measure the dependence of star formation on cluster mass
for the high-redshift sample.  However,
because the SDSS and 3 \edi \ clusters presented here span the same range of 
velocity dispersions,
the zeropoint calculation will yield a similar number even if the 
slope of the relation changes with redshift.
Thus our measure of the average evolution in the total \ha \ luminosity 
between $z = 0.07$ and $z \simeq 0.75$ should be insensitive to changes in
the slope of the $\Sigma L(H\alpha) - \sigma$ relation, although the interpretation
of this result could be different.

\begin{figure}[h]
\plotone{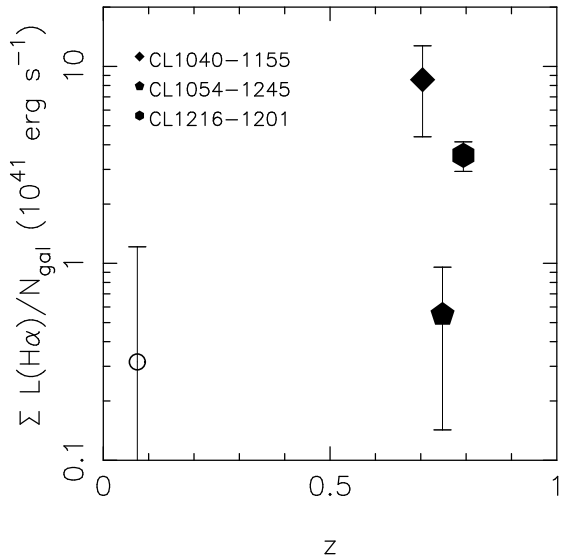}
\caption{Integrated \ha \ luminosity divided by number of member galaxies versus cluster redshift
for C4 clusters (open circle) and $z\simeq 0.75$ \edi \ clusters (filled symbols).  The point for the
C4 clusters shows the average for the 308 clusters, and the errobars
show the standard deviation.
\label{hangalz_f10}}
\end{figure}

\begin{figure}[h]
\plotone{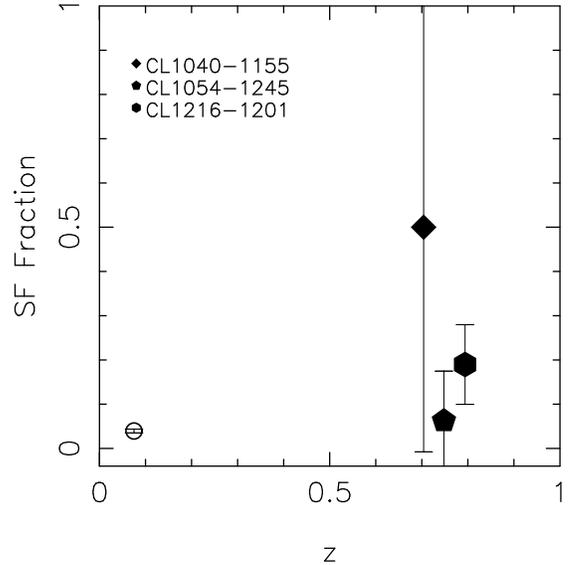}
\caption{Fraction of star-forming galaxies versus cluster redshift
for C4 clusters (open circle) and $z\simeq 0.75$ \edi \ clusters (filled symbols).
\label{hangalz_f11}}
\end{figure}

As with the analysis in \S\ref{intcluster}, we try to reduce the dependence 
of the dependent variable, the 
integrated \ha \ luminosity, on $\sigma$ by normalizing by the total number of member galaxies.  
For the \edi \ clusters, we use the estimate of 
the total number of galaxies that is derived from the luminosity functions of {Rudnick} {et~al.} (2007); 
these estimates have associated uncertainties of 30-50\%.  
The results, shown
in Figure \ref{hangalz_f10}, indicate that the average \ha \ luminosity per galaxy has 
increased by a factor of $13 \pm 7$
between $z \simeq 0.07$ and $z \simeq 0.75$.  
We show the fraction of star-forming galaxies versus redshift in Figure \ref{hangalz_f11}.
The fraction of star-forming galaxies (where star-forming galaxies are 
those that meet the selection criteria outlined above) 
is a factor of $6\pm 3$ higher in the \edi \ clusters.
Poggianti et al. (2006) find that the fraction of star-forming galaxies
varies with cluster mass at higher redshift.  
Again, we are comparing with only 3 higher-redshift clusters, and so we do not
attempt to quantify a mass-dependent evolution in the star-forming fraction.

\subsection{Distribution of \ha \ Luminosities \label{distlha}}
\begin{figure*}
\plotone{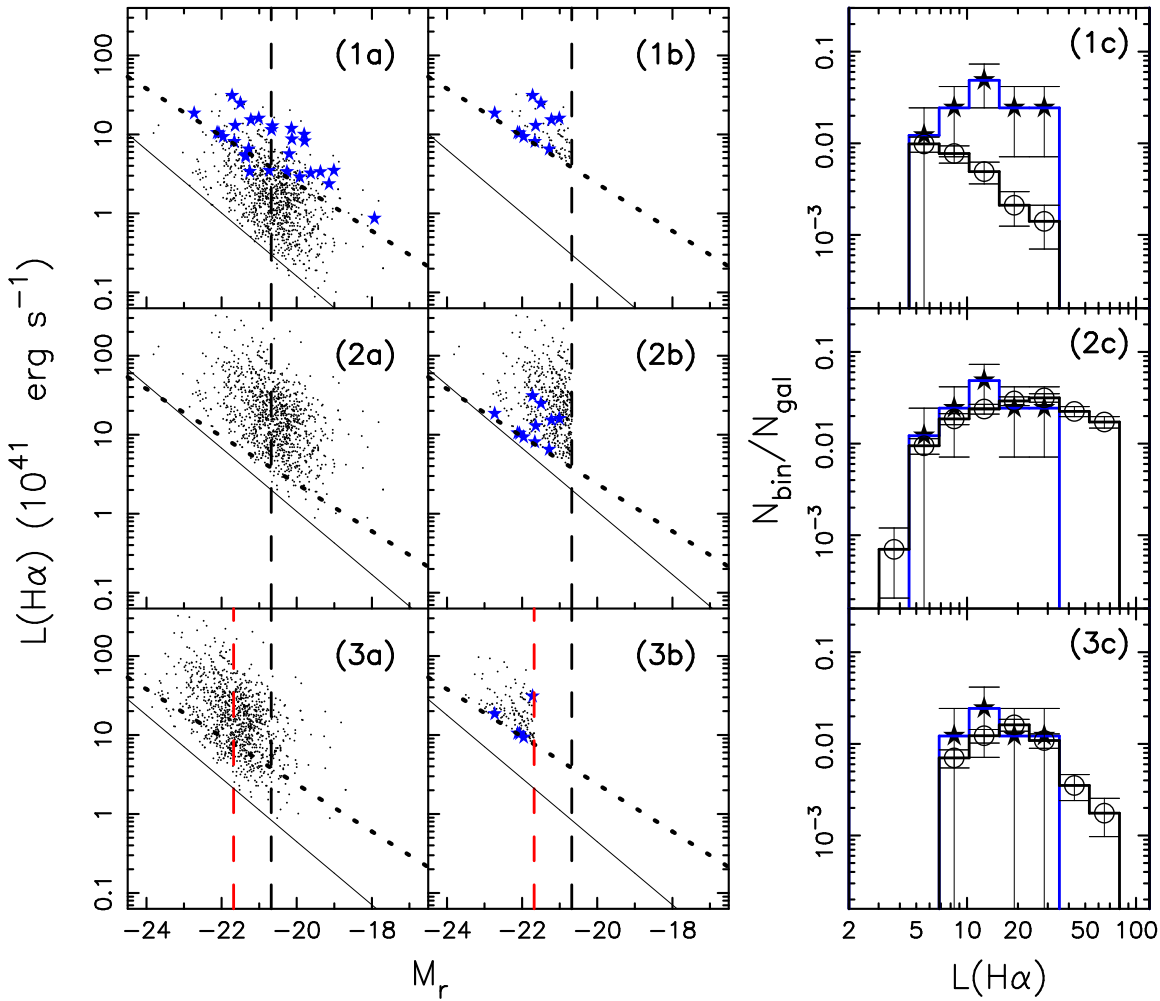}
\caption{{\bf (1a)} L(H$\alpha$) versus M$_r$ for SDSS galaxies from 308 clusters 
(black dots) and $z\simeq 0.75$ galaxies (blue stars) from 3 clusters.  Vertical
dashed line shows spectroscopic completeness limit of SDSS.  The solid line shows the EW limit
of the SDSS spectra.  The dotted line shows flux limit
of $z \simeq 0.75$ surveys.  
{\bf (1b)} Same as panel 1a but showing only those galaxies that lie above the 
flux and magnitude limits.  
{\bf (1c)}  Distribution of \lha \ for galaxies in panel 1b.
{\bf (2a)} L(H$\alpha$) versus M$_r$ for SDSS galaxies after \lha \ is scaled by a factor of 10.
{\bf (2b)} Same as panel 2a but showing only those galaxies that lie above the flux and magnitude limits.
{\bf (2c)} Distribution of \lha \ for galaxies in panel 2b.
{\bf (3a)} L(H$\alpha$) versus M$_r$ for SDSS galaxies after \lha \ is scaled by a factor of 3 and
M$_r$ is brightened by 1 magnitude.  The red dashed line shows the magnitude of the SDSS spectroscopic completeness
limit, brightened by 1 magnitude. 
{\bf (3b)} Same as 3a but showing showing only those galaxies that lie above the flux and magnitude limits. 
{\bf (3c)} Distribution of \lha \ for galaxies in panel 3b.
\label{sfrmabsmulti}}
\end{figure*}

We now try to understand the global evolution of cluster star-formation 
properties in terms of the evolution of individual star-forming galaxies.  
We show the low and high-redshift star-forming galaxies in Figure \ref{sfrmabsmulti}, where 
we plot \lha \ versus M$_r$ in panel (1a) for
all galaxies within
0.5$\times$\rtwo \ and $\Delta v < 3 \sigma$.  
The black points show the galaxies in the 308 SDSS clusters and the filled stars
show the galaxies in the three $z\simeq 0.75$ clusters (we do not correct for the
additional field contamination in \ca \ and \cb).  
The dotted line in Figure \ref{sfrmabsmulti} shows the approximate flux 
limit of the $z\simeq 0.75$ imaging, and the 
dashed vertical line shows the magnitude corresponding to the SDSS spectroscopic 
completeness limit.
The solid gray line shows the approximate effect of imposing a minimum 
\ha \ EW cut of 4\AA \ for the SDSS galaxies.
The galaxies that meet the high-redshift flux limit and the SDSS spectroscopic magnitude cut are
shown in panel (1b) of Figure \ref{sfrmabsmulti}.

To compare the distribution of \ha \ luminosities, 
we consider only the star-forming galaxies that lie above the 
high-redshift flux limit and are brighter than the SDSS spectroscopic limit 
(shown in panel (1b), Figure \ref{sfrmabsmulti}) so that
we are comparing relatively complete populations in both cluster samples.  In panel (1c) we show the
number of galaxies per bin divided by the total number of cluster members 
versus \lha \ for the SDSS galaxies (open circles) and \edi \ 
galaxies (filled stars).  The high-redshift clusters contain a higher fraction of the star-forming galaxies.  
This could be explained by pure number evolution, if the fraction of
star-forming galaxies per cluster increased by a factor of $\sim10$,
with no change in the distribution of \ha \ luminosities of star-forming galaxies.  
However, the SDSS \lha \ distribution peaks at a lower value of \lha \ than the \edi \ distribution, 
indicating that there is some evolution of the \ha \ luminosities of individual galaxies.

As an example of evolution that is due solely to an increase in the
average star-formation rate per galaxy (among the active population), 
we simply scale the \ha \ luminosities of individual SDSS galaxies by a factor of 10, which is the
amount of evolution observed in the integrated \ha \ luminosity.  
We show the scaled SDSS galaxies in panel (2a) of Figure \ref{sfrmabsmulti}, the comparable samples of 
SDSS and \edi \ galaxies in panel (2b), and 
the corresponding distribution of \ha \ luminosities in panel (2c).  
This simple scaling brings the number of star-forming
galaxies in the SDSS clusters 
into agreement with the \edi \ distribution.
Note that because we have observed only three high-redshift clusters, we do not expect to detect high-redshift
galaxies in bins where the expected counts fall below 0.33 galaxies
per cluster.  

While this simple model does an excellent job of reproducing the 
evolution of the distribution of \ha \ luminosities 
and is consistent with the evolution of the integrated \ha \ luminosity, 
we must also consider that the broad-band magnitudes of galaxies evolve.
On average, galaxies have faded since $z\simeq 0.75$, and this impacts the inferred \lha \ distribution
because galaxies in the high-redshift sample may have faded below the magnitude cut in the low-redshift
sample.  We account for this in a second model by 
brightening each SDSS galaxy by 1 magnitude to account for fading since $z \simeq 0.75$ 
(e.g. {Poggianti} \& {Barbaro} 1997) while simultaneously 
scaling the \ha \ luminosities of SDSS galaxies by a factor
of 3.  The scaled 
SDSS galaxies are shown in panel (3a), and the galaxies that 
meet the \edi \  flux cut and the SDSS magnitude cut (which is now a magnitude 
brighter and shown with the red dashed
vertical line) 
are shown in panel (3b).  The resulting \lha \ distributions (shown
in panel (3c)) are in reasonable agreement, although the sample of \edi \ galaxies 
used to derive the $z \simeq 0.75$ \lha \ distribution is 
uncomfortably small.  

These models show that the evolution of the star-forming cluster galaxies between
$z\simeq 0.75$ and $z \simeq 0.07$ can be 
characterized by a fading of both the \ha \ and broad-band luminosities of individual galaxies.  
However, the two scenarios we present in Figure \ref{sfrmabsmulti} are by no
means unique, and further study is required to characterize the evolution of the \ha \ luminosity 
distribution
in terms of the evolution of the number, magnitudes, and star-formation rates of the star-forming galaxies.
Nonetheless, these preliminary results
suggest that the \ha \ luminosities of individual cluster galaxies 
have faded by a factor of up to $\sim10$ since $z\simeq 0.75$.  
This factor of $\sim 10$ is an upper limit
on the evolution, and it will be lower if there is evolution of galaxy magnitudes.  In
addition, galaxies that are presently on the red sequence may have been forming stars at
$z \simeq 0.75$ (e.g. {Bell} {et~al.} 2007), and this may further decrease 
the inferred evolution of individual \ha \ luminosities.

Several issues compromise our comparison of the SDSS and $z\simeq 0.75$ star-forming 
galaxies.  
The dominant limitation is the small number 
of high-redshift clusters.  We have shown in \S\ref{results} that 
$\rm \Sigma$SFR exhibits large scatter as a function of cluster mass, so a more
precise measure of evolution requires a much larger sample of high-redshift clusters.
A second limitation is our use of \ha\ as a star-formation indicator.
Although relatively insensitive to dust compared with other commonly
used optical indicators such as [OII] emission
(e.g. {Nakata} {et~al.} 2005), it is well known that a full census of star
formation must account for energy re-radiated by dust
(e.g. {Duc} {et~al.} 2002; {Coia} {et~al.} 2005; {Bell} {et~al.} 2007; {Bai} {et~al.} 2007).  If dust obscuration is a
strong function of redshift and environment, this could affect our
conclusions.  Furthermore, for the high-redshift clusters
we are not able to quantify the AGN contamination through the \ha\
emission alone.
AGN account for an average (median) of  23\%(34\%) of total \ha \ emission 
for galaxies within 0.5\rtwo \ and $\Delta v < 3\sigma$ for the SDSS clusters 
and 20\% (average) by number.
Existing studies of high-redshift clusters show low AGN fractions 
(e.g., {Homeier} {et~al.} 2005), so AGN contamination is probably not a limiting factor
in the comparison.  
Upcoming studies of the high-redshift clusters 
at X-ray and infrared wavelengths will 
provide some insight into the AGN fraction of the \edi \ clusters.
A final caveat that may affect our findings in \S\ref{results} and \S\ref{evolution} 
is that the SDSS \ha \ luminosities are derived from 
aperture fluxes. {Koopmann} {et~al.} (2006) find radial gradients in \ha \ emission that are not
reflected in the radial distribution of stellar light, and this compromises the aperture
correction we apply to the \ha \ flux.  As a result, 
our analysis is not sensitive to processes that preferentially affect the perimeter of galaxies.

\subsection{Comparison with Field Evolution}
Field surveys show a factor of $\sim$10 decrease in $\rm \Sigma$SFR/Mpc$^3$ since $z \sim 0.8$ 
(e.g. {Tresse} \& {Maddox} 1998; {Glazebrook} {et~al.} 1999; {Tresse} {et~al.} 2002; {Hippelein} {et~al.} 2003; {Bell} {et~al.} 2007).  
In comparison,  
Figure \ref{sfrmabsmulti} shows that we can reproduce the evolution between the
 star-forming galaxies in the low and high-redshift clusters by scaling the \ha \ luminosity of
the SDSS galaxies by a factor of $\sim10$.  This is remarkably
similar to what is seen in the field, and suggests that the strong evolution
observed is not driven by cluster-specific physics.  However,
it is important to distinguish 
the average star-formation rate per galaxy 
and the average star-formation rate of the star-forming galaxies. 
Recent work by {Bell} {et~al.} (2007), based on the COMBO-17 photometric
redshift survey, suggests that the field evolution is due
in roughly equal parts to an increase in the number density of
star-forming galaxies, and in the average star-formation rate per
galaxy among this population.  
Our analysis of the \ha \ luminosity distribution suggests a similar
division in clusters, with an increase in galaxy star-formation rates
contributing about half of the factor of $\sim 10$ observed evolution in
the integrated \ha \ luminosity.  However, we require a larger
sample of high-redshift clusters, preferably with complete redshifts, and
a comparable sample of high-redshift field galaxies to confirm this.

An issue that complicates the comparison in evolutionary rates
between clusters and the general field is
that clusters have approximately doubled
their mass over the time range we are probing.  
However, many of the accreted
galaxies may already be in the high-redshift survey volume (the $\pm 3 \sigma$ velocity
cut corresponds to a large volume).  
A careful comparison with simulations is needed to understand 
how infalling galaxies affect our measure of cluster star-formation properties.

\subsection{Scenarios for Declining SFRs}
Our main result is that the rate of evolution of clusters does not appear to be significantly
different from that of the field. 
This suggests the cluster environment has not had a direct effect on
galaxy star-formation rates over the past few billion years.  
Either the difference we observe between cluster and field populations
was imprinted at high redshift, or it is driven by physics occurring on
group scales (or both).  
Here we discuss the implications of our results on the various mechanisms frequently
used to explain the decline of cluster star-formation rates.   We remind the reader
that our observations are restricted to relatively massive galaxies,
and lower mass galaxies may behave quite differently.  

If dense environments are to directly influence star-formation rates at
low redshift ($z<1$), they must presumably do so by removing gas from
galaxies, since this is the dominant reservoir of baryons.  Typically,
discussion of such processes is divided into three types: the
rapid consumption of gas, through massive starbursts; the removal of
{\it cold} gas bound to the disk, via ram-pressure stripping ({Gunn} \& {Gott} 1972)
or other thermal mechanisms ({Nulsen} 1982); and the removal of {\it
  hot}, more loosely bound gas, through similar mechanisms ({Larson} {et~al.} 1980; {Kawata} \& {Mulchaey} 2007).
All of these mechanisms will lead to a reduction of star formation, but
they operate on different timescales and in different environments.
The efficacy of ram pressure stripping
depends on the velocity of the galaxy relative the ICM and the density of the ICM.  Both
factors reach their maximum values at the cluster center, and so ram-pressure stripping should 
be most effective for galaxies plunging through the middle of the cluster.  One also expects
that ram-pressure stripping is more effective in higher-mass clusters because they have 
higher-density ICMs (e.g. {Mohr} {et~al.} 2000; {Vikhlinin} {et~al.} 2006), and their
velocity dispersions are higher.  It is actually difficult to strip away
{\it all} the cold gas in a disk, except in the centers of the most
massive clusters (e.g. {Quilis} {et~al.} 2000; {Roediger} \& {Brueggen} 2007).  The fact that we see
no trends in average star-formation rate as a function of cluster
mass suggests\footnote{
A possible caveat is that, as noted above, {Koopmann} {et~al.} (2006) show that star-formation is truncated 
at the outer edge of
cluster spirals in Virgo, and the SDSS fiber-derived \ha \ fluxes are not able to detect processes that
preferentially affect the outer radii of galaxies.} that ram-pressure stripping of cold gas is 
not the dominant mechanism driving the 
evolution of star-forming cluster galaxies although it certainly
occurs in some cases ({Kenney} {et~al.} 2004; {Koopmann} {et~al.} 2006).  Instead, the 
decline in the star-formation rates of cluster galaxies 
since $z \sim 1$ must be primarily driven by physical processes that do not depend
on cluster mass, at least for clusters with velocity dispersions
greater than 450~km~s$^{-1}$.   A similar argument holds for the consumption
of gas by starbursts, since such bursts are most easily caused by
galaxy mergers which should be rare in cluster environments (but could
be common in the outskirts, or in groups).

In hierarchical clustering models, clusters 
form in regions of high overdensity and grow through the
accretion of field galaxies and groups.
Most models of galaxy formation include a crude treatment of
environmental effects, where satellite galaxies of a halo of any mass
are instantaneously stripped of their hot gas, preventing further growth
of the cold gas reservoir (e.g., {White} \& {Rees} 1978; {Kauffmann} 1995; {Bower} {et~al.} 2006; {Croton} {et~al.} 2006; {De Lucia} {et~al.} 2006).  
This "starvation" mechanism (originally proposed by {Larson} {et~al.}  1980)
serves to shut down star formation on $\sim$1-2 Gyr
timescales, starting from the instant a galaxy becomes a satellite in a
larger halo. Undoubtedly this gas is easier to remove than the cold gas
considered above ({Kawata} \& {Mulchaey} 2007); however, it is becoming increasingly clear that the
simple assumption of instant removal of gas on any satellite,
regardless of halo mass, is 
too effective at suppressing star formation (e.g. {Poggianti} {et~al.} 2006; {Weinmann} {et~al.} 2006b; {McCarthy} {et~al.} 2007).
Our results also suggest that star-formation is not shut off immediately;
rather, we see a relatively gradual fading in \lha\ of a factor $\sim 3-10$ over a timespan of $\sim$7 billion years.
Starvation that operates over longer timescales can also explain
why field galaxies evolve at a similar rate, where the decline in field star-formation rates 
is due to the build-up of groups.

\begin{figure}[h]
\plotone{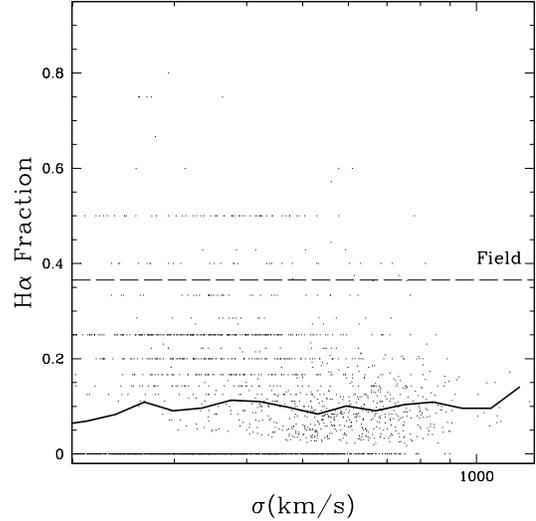}
\caption{The predicted fraction of \ha-emitting galaxies versus cluster velocity
  dispersion for $z=0$ clusters selected from the model of
  {Bower} {et~al.} (2006).  Each cluster is shown as an individual dot.  The
  many clusters with no emission-line galaxies are represented with a
  mean of 0.05 and artificial scatter, for visualization purposes only.
  The galaxy magnitude limit, and \ha\ flux and EW limits, are the same
  as in our SDSS sample.  The horizontal dotted line represents the average
  fraction among all $z=0$ galaxies in the model (above the same
  magnitude limits).  The thick solid line represents the average
  emission line fraction in bins of velocity dispersion.  
\label{hasim}}
\end{figure}

Another way to see this is to directly compare the amount of star
formation in our clusters with predictions from the models.  For this
purpose, we use the publicly available {Bower} {et~al.} (2006) models, based on
the Millennium simulation.  For each parent dark matter halo we
calculate the expected value of $\sigma$ and $R_{200}$ using the same
relations used here (Eqn. \ref{eqn-r200}), and select galaxies based on their
projected position, relative velocity (including Hubble flow and
peculiar velocity components) and $r$-band luminosity.  Using this
membership criteria we compute the fraction of galaxies with H$\alpha$
emission satisfying our observational criteria on flux and EW, as a
function of velocity dispersion.  The results are shown in
Figure~\ref{hasim}.  As in the observations, there is no significant
trend with velocity dispersion.  However the average fraction of galaxies with H$\alpha$ emission
is only 10\%, significantly lower than our observed value of 20\% (Figure \ref{allvsigma}).
This difference alone could simply be reflecting uncertainties in
modeling H$\alpha$ emission, which depends on the dust model,
metallicity, and AGN contribution.  But interestingly the average
emission line fraction for the model {\it field} population is 36.5\%,
in very good agreement with our control sample value.  Thus, the
difference in emission-line fraction in clusters relative to the field
is significantly larger in the models than in our data, and this should
be robust to modeling of \lha\ (but of course sensitive to environmental
dust or metallicity effects).  This lends support to
our suggestion that the truncation of star formation in clusters is too
severe in the current models.  Note that the models assume a
suppression of star formation that is not only rapid but also efficient
in low mass systems, well below the mass range our cluster sample.  This
is because the starvation model as implemented has little dependence on
the mass of the host halo, which is probably unrealistic
({McCarthy} {et~al.} 2007).  Recent observations suggest that groups have
substantially {\it higher} star-formation rates than clusters
(e.g. {Wilman} {et~al.} 2005; {Balogh} {et~al.} 2007), in contrast to these simple model
predictions, and therefore these low mass systems are an important
environment worthy of further study.

\section{Summary \& Conclusions}
\label{conclusions}

Using 308 dynamically relaxed, low-redshift clusters from the C4 cluster catalog, we 
investigate the dependence of the total stellar mass, 
number of star-forming galaxies, and total star-formation rate on 
cluster velocity dispersion.
We find that the total stellar mass, the number of star-forming galaxies, and total star-formation rate
scale linearly with the number of galaxies with no dependence on velocity dispersion.  
In other words,
$\rm \Sigma L(H\alpha)/N_{gal}$ and $\rm N_{SF}/N_{gal}$ are
independent of velocity dispersion.
We interpret this to mean that the cluster environment at low redshift does not
differentially affect the star-forming properties of luminous galaxies
($\rm M_r < -20.68$) across dense
environments characterized by $\sigma > 450$~km~s$^{-1}$.

We compare the SDSS clusters with a sample of $z\simeq 0.75$ clusters from 
the literature and find that the star-forming cluster population has 
declined significantly since $z\simeq 0.75$.  
Using the relationship between $\rm \Sigma L_{H\alpha} - \sigma$ 
defined from the local cluster sample, we measure
the mean shift in this relationship to the high-redshift cluster sample. On
average
(correcting for the mass growth of clusters between the two redshifts)
we find that the total \ha \ luminosity of the high-redshift clusters 
is $\sim10$ times greater than that of the low-redshift clusters.  In more detail,
we find that we 
are able to bring the SDSS \ha \ luminosity distribution into agreement
with that of the high-redshift clusters by scaling the \lha \ of individual SDSS galaxies
by up to a factor of up to $\sim 10$, and we can reduce the required evolution in 
\ha \ luminosities to a factor of $\sim 3$ if we allow for fading of galaxies between
$z \simeq 0.75 $ and $z = 0.07$.  
Thus, using the SDSS clusters, we are able to break 
the degeneracy between redshift and mass-evolution that limited our
previous attempts to quantify the evolution of cluster star-formation 
properties ({Finn} {et~al.} 2004, 2005).

We compare the cluster results to the evolution seen in field galaxies 
over a similar redshift interval.
We find a factor of $\sim$10 decline in total star formation with decreasing redshift
for the clusters as compared to a factor of $\sim$10 decrease for the field. 
Thus, we find no evidence for any differential evolution 
between cluster and field galaxies over this redshift
range.  This strengthens our conclusion that the effect of the 
cluster ($\sigma > 450$~km~s$^{-1}$)
environment on bright, star-forming
galaxies is relatively modest for $0 < z < 0.9$.
In addition, the mechanism causing the decline in galaxy star-formation rates 
must operate over a fairly long timescale 
such that the star-formation rates of individual galaxies decline
by an order of magnitude over $\sim$7 billion years, a decline timescale which is longer
than galaxy evolution models commonly assume.

\acknowledgements
RAF thanks Greg Rudnick, Bianca Poggianti, Houjun Mo, Xiaohu Yang, Daniel McIntosh, 
Martin Weinberg and Ken Rines 
for useful discussions regarding this work.  
RAF acknowledges support from an NSF Astronomy and Astrophysics
Postdoctoral Fellowship under award AST-0301328.  
MLB gratefully acknowledges support from an NSERC
Discovery grant.  DZ acknowledges financial support for this work from a Guggenheim
fellowship and NASA
LTSA award NNG05GE82G, and thanks the NYU Physics department and Center for
Cosmology and Particle Physics for support and hospitality during his
sabbatical.

This research has made use of  
NASA's Astrophysics Data System.
Funding for the Sloan Digital Sky Survey (SDSS) and SDSS-II has been provided by the 
Alfred P. Sloan Foundation, the Participating Institutions, the National Science Foundation, 
the U.S. Department of Energy, the National Aeronautics and Space Administration, the Japanese 
Monbukagakusho, and the Max Planck Society, and the Higher Education Funding Council for England. 
The SDSS Web site is http://www.sdss.org/.

\appendix
To estimate the additional field contamination in our high redshift
surveys, due to the large velocity range sampled by the narrow-band filters, 
we use the mock redshift survey of 
{Yang} {et~al.} (2004).  We use all halos with masses greater than 
$\rm 10^{14}~M_\odot$, which corresponds roughly to the mass threshold imposed 
by our $rm \sigma > 450~km~s^{-1}$ cut, and 
select all galaxies within $0.5 \times R_{vir}$ and $M_{B_J} < -18$.  Here, Yang 
{et~al.} use $R_{180}$ to define the virialized region of each halo; this agrees 
to within 5\% with \rtwo (the radial cut we have used throughout), 
and this slightly different radial cut does not 
affect the inferred contamination.  We
define member galaxies as those physically located within the virial radius and
contamination as the number of non-members normalized by the number of observed
galaxies.  We calculate the contamination for late-type galaxies, which are 
likely to be the ones contributing to \ha \ emission and which also have a higher
contamination rate than early-type galaxies.  
The simulation size is 300~$h^{-1}$~Mpc, and the largest velocity dispersion is 2100~km/s.  
A $\pm 6 \sigma$ velocity cut corresponds to a volume of 252~$h^{-1}$~Mpc, so 
we are not limited by edge effects.  
We find that the contamination of late-type galaxies is 25\% in a $\pm 3 \sigma$ velocity cut and 31\% in 
$6 \sigma$ cut.  Therefore, we correct the integrated \lha \ of \ca \ and \cb \ by
a factor of $31/25 = 0.81$ to account for additional field contamination.
The {Yang} {et~al.} (2004) simulation is not 
an ideal comparison 
because it is at a redshift of zero, has B-band magnitudes, and is normalized to the
2dFGRS survey.  Nonetheless, it gives 
a useable estimate of the extra field contamination for \ca \ and \cb.  


\end{document}